\documentstyle[12pt]{article}

\def\beq{\begin{equation}}   \def\eeq{\end{equation}}

\newcommand{\gsim}{\lower.7ex\hbox{$\;\stackrel{\textstyle>}{\sim}\;$}}
\newcommand{\lsim}{\lower.7ex\hbox{$\;\stackrel{\textstyle<}{\sim}\;$}}

\newcommand{\ra}{\rightarrow}

\begin{document}

\def\lsim{\mathrel{\rlap{\lower3pt\hbox{\hskip0pt$\sim$}}
    \raise1pt\hbox{$<$}}}         
\def\gsim{\mathrel{\rlap{\lower4pt\hbox{\hskip1pt$\sim$}}
    \raise1pt\hbox{$>$}}}         

\begin{titlepage}
\renewcommand{\thefootnote}{\fnsymbol{footnote}}

\begin{flushright}
CERN-TH/96-335\\
TPI-MINN-96/21-T\\
UMN-TH-1519/96\\
hep-th/9611213\\

\end{flushright}

\vspace{0.3cm}

\begin{center}
\baselineskip25pt

{\Large\bf 
Dynamical Compactification as a Mechanism of Spontaneous Supersymmetry 
Breaking}

\end{center}

\vspace{0.3cm}

\begin{center}
\baselineskip12pt

\def\thefootnote{\fnsymbol{footnote}}

{\large G. Dvali}

\vspace{0.1cm}
Theory Division, CERN, CH-1211 Geneva 23, Switzerland
\vspace{0.2cm}

{\em and}

\vspace{0.3cm}
{\large  M.~Shifman} 

\vspace{0.1cm}
Theory Division, CERN, CH-1211 Geneva 23, Switzerland \\

and \\

  Theoretical Physics Institute, University of Minnesota, Minneapolis,
MN 54555 USA$^\dagger$ \\[0.5cm]

\vspace{0.7cm}

{\large\bf Abstract} \vspace*{.25cm}
\end{center}

Supersymmetry breaking
and compactification of  extra space-time dimensions may
have a common dynamical origin if our Universe is
spontaneously generated in the form of  a  four-dimensional topological
or non-topological defect in higher-dimensional space-time.
Within such an approach  the conventional particles are zero modes
trapped in the core of the defect. In many cases  solutions of this type 
{\em spontaneously} break all supersymmetries of the original theory,  so that 
the low-energy observer from ``our" Universe
inside the core would not detect supersymmetry. Since the extra dimensions
are not compact but, rather,  inaccessible to low-energy observers,
the usual infinite tower of the Kaluza-Klein excitations does not exist.
Production of superpartners at the energy scale of SUSY restoration will be 
accompanied by  four-momentum non-conservation. (Depending on the
nature of the solution at hand, the non-conservation may either happen above
some threshold energy or be continuous). In either case, the door to extra 
dimensions may be not very far from the energies accessible
at  present colliders.

\vspace{0.3cm}

\hfill

\begin{flushleft}
CERN--TH/96--335\\

November 1996

\vspace{0.25cm}

\rule{2.4in}{.25mm} \\
$^\dagger$ Permanent address.

\end{flushleft}

\end{titlepage}

\section{Introduction}

The old idea of Kaluza and Klein \cite{kaluza} that our Universe may have 
more than four
dimensions is a basic element of most of the modern particle physics theories,
especially of those that are based on supergravity or superstrings. Two 
central problems of these theories are: 1) hiding the additional space-time
dimensions and 2) the origin of the supersymmetry (SUSY) breaking.
The solution of the first problem is usually
attributed to a mechanism of ``compactification", which assumes that extra
dimensions are compact, with a small radius, and, thus, are invisible for the
four-dimensional observer (for a review see \cite{Appe}).
An alternative attractive idea, to be explored in this paper,
is that our Universe can be a four-dimensional topological defect 
dynamically occurring in a
higher-dimensional space-time.
Ordinary particles in this approach can be viewed as ``zero modes" trapped in
the core of the defect. 
We will refer to this mechanism as {\em dynamical compactification}.
The defect need not be
 topological; other locally stable field configurations can do
the same job just as well. A key difference of this approach from the 
conventional
spontaneous
compactification schemes \cite{SC} is that extra dimensions are not compact, 
but are
invisible simply because the zero modes cannot escape from the core of the
defect (into the region where they gain large mass). 
In this sense the
four-dimensional Universe is {\em dynamically} generated out of a higher
dimensional one, through a spontaneous breaking of (a part of) the 
translational
symmetries of the original theory, which in particular result from
a cosmological phase transition with spontaneous (internal) symmetry 
breaking.
A nice example of this phenomenon is 
provided by a domain 
wall. This particular example was suggested almost 15 years ago
in Ref.
\cite{RS}, where it was shown that the extra dimensions will be hidden
in this case, much in the same way as in the conventional
spontaneous
compactification
with no translational symmetry breaking \cite{SC}, 
provided that the excitation energies of the order of the masses of the 
non-zero 
modes are inaccessible.

As was just mentioned above,  another yet unresolved issue is the question of
supersymmetry breaking. Within the standard approach this question
is not directly related to the compactification scheme 
\footnote{The only exceptions we know of  are the so-called 
``coordinate-dependent 
compactification" 
\cite{ss,rohm} and the introduction of a constant magnetic field in the compact 
dimensions
\cite{magnet}. Both mechanisms are formulated directly at the string level. In 
the present paper we confine ourselves to the  field theory consideration
of phenomena  at energies that are small in the Planck scale.}. The key 
observation of 
the present work is the fact that, when
implemented in a supersymmetric context, the idea of dimensional reduction 
through the dynamical compactification
leads to a spontaneous supersymmetry breaking, with a very peculiar
pattern. This is a new mechanism which, to our knowledge,
was not discussed previously, and which 
 may shed  light on the question
why we do not see any Fermi-Bose symmetry in our Universe. If the 
conventional compactification mechanisms are usually associated with the 
Planck radius, the mechanism to be discussed below can occur at any scale.
The scale of the supersymmetry breaking is then related to the scale
of the dynamical compactification. The fact that the extra dimensions
in the dynamical compactification may be non-compact
leads to a remarkable prediction: once we cross a threshold
allowing one to excite (non-localized) non-zero modes,
the four-dimensional observer will effectively  see a strong non-conservation 
of  four-momentum \cite{RS}. The above mentioned threshold is obviously
of the order of the scale
of the supersymmetry restoration. Thus, if this mechanism is relevant to 
nature, at a certain scale we will simultaneously discover 
superpartners and non-conservation of  four-momentum. 

As  toy examples, we first  consider embedding of the $(2 + 1)-$ and 
$(1 + 1)$-dimensional
worlds into the four-dimensional Universe in the form of a domain wall
and  a cosmic string, respectively. 
These objects, obviously, spontaneously break (a part of) the original 
translational invariance. In the full four-dimensional theory,
the full translational invariance is restored since for every given wall
there is another one, shifted in a certain way.  If we are bound to live
inside a given wall, however, the  translational invariance is partly broken.
This naturally results in the spontaneous breaking of (a part) of 
supersymmetry. The Goldstone excitations corresponding to the broken
bosonic and fermionic generators are the zero modes trapped in the core of 
the defect. Sometimes the zero modes occur for reasons other than the 
Goldstone theorem. 
``Quarks" and ``leptons" are identified
with these  zero modes. In the simplest model, with one complex scalar field,
the dynamical compactification through the domain wall breaks a half of 
supersymmetry.

Although it  often happens  
that only  a half of  supersymmetry is broken (this phenomenon is not new
\cite{wittenolive, Bag}), 
other examples, where SUSY is 
completely broken, are also abundant. We will mainly focus on these 
examples. As a matter of fact, leaving the domain 
wall intact, but adding extra
particles to the original theory, we arrive at a  
model, with the domain wall, where all SUSY generators are spontaneously
broken inside the wall. Another example where the complete SUSY breaking 
takes place is the dynamical  compactification from
3+1 into 1+1 inside a vortex.

This is not the end of the story, however.
We find a similar effect generated by $(2 + 1)$- and
$(1 + 1)$-dimensional defects that are not domain walls or strings in the
usual sense. 
New opportunities are provided by the fact that in some SUSY theories
the vacuum is continuously degenerate, i.e. there exist flat directions. 
This allows one to
find stable field configurations (breaking the translational 
invariance) characterized by the spread out of the Higgs field
everywhere on the 
 vacuum manifold (which
has a non-compact flat direction).
The energy density is purely gradient. Such new solutions break the
full supersymmetry of the initial theory. As a result, in  both cases
the dimensionally reduced theory looks as non-supersymmetric.
This mechanism of spontaneous SUSY breaking has  peculiarities,
to be discussed below. In particular, the threshold for the momentum 
non-conservation is absent.

Thus, we found  a rich spectrum of various scenarios
of  spontaneous SUSY breaking associated with the dynamical 
compactifications. Whether any of these scenarios can be exploited
in the context of realistic, phenomenologically successful models
remains an open question; in the present paper we will sketch only one, 
rather obvious, scheme. 

The paper is organized as follows. In Sect. 2 we first briefly review general
ideas associated with the dynamical compactification.
Then, starting from a known example, we derive a criterion
allowing one to tell in which cases the wall-like defects will break 
supersymmetry completely (Sect. 2.2). In Sect. 2.3 an explicit model of this
type, with a non-minimal wall, is constructed.  Sect. 3 treats the dynamical
compactification on global cosmic strings.  In Sect. 4 we gauge the global 
$U(1)$ 
symmetry and consider a gauge string example. In Sect. 5 a new class of 
solutions is introduced, which become possible thanks to the
existence of non-compact flat directions in some SUSY theories.
Finally, Sect. 6 treats SUSY breaking through a global winding.
This  mechanism combines some features of the Kaluza-Klein approach
with the dynamical compactification. In Sect. 7 a possible
phenomenological application is outlined. 

\section{Dynamical  Compactification on  Domain Wall}

We start from the  simplest globally supersymmetric system,
which admits a topologically stable (domain wall) solution in  four
dimensions, the  Wess-Zumino model \cite{WZ}
with one complex scalar
$\Phi$ and its fermionic superpartner, a four-component real (Majorana) 
spinor
$\psi$. These two fields compose a minimal irreducible-representation
chiral superfield $\Phi$. Here and below  we  denote the chiral superfields
and their scalar components by one and  the same letter; in each particular
case it will be clear from the context which one we refer to. 

In the component notations the Lagrangian of the Wess-Zumino
model has the form
$$
L= {\partial^\mu \Phi^\dagger}{\partial_\mu\Phi}
+\frac{1}{2}\bar\psi\gamma^\mu\partial_\mu\psi
 + F^\dagger F +
$$
\begin{equation}
\left\{ \left[ \lambda\mu^2 F 
-\lambda
\left(\Phi^2F+ \Phi 
\bar\psi ({1+i\gamma^5 \over 2})\psi\right)  \right] +
{\rm h.c.} \, \right\}
\label{LWZM}
\end{equation}
where the superpotential is conveniently chosen as 
\begin{equation}
W = \lambda(-\Phi^3/3 + \mu^2 \Phi)\, .
\label{superpot}
\end{equation}
It exhibits a discrete $R$ symmetry $\Phi \rightarrow -\Phi$, 
spontaneously broken by the vacuum expectation value (VEV) $\Phi = \pm 
\mu$. Our $\gamma$ matrices are all real (they differ from the
standard Majorana matrices by $i$),
$$
\gamma^0 = 
\left(
\begin{array}{cc}
0 & -\sigma_1 \\
\sigma_1 & 0 
\end{array}
\right)\, ,\,\,\,
\gamma^1 = 
\left(
\begin{array}{cc}
0 & -1 \\
-1 & 0 
\end{array}
\right)\, ,\,\,\,
\gamma^2 = 
\left(
\begin{array}{cc}
0 & -i\sigma_2 \\
i\sigma_2 & 0 
\end{array}
\right)\, ,\,\,\,
\gamma^3 = 
\left(
\begin{array}{cc}
1 & 0 \\
0 & -1
\end{array}
\right)\, ,
$$
\beq
\gamma^5= 
\left(
\begin{array}{cc}
0 & -\sigma_3 \\
\sigma_3 & 0 
\end{array}
\right)\, .
\label{gamma}
\eeq

As  is well known, this system admits a topologically stable kink solution
independent of the three coordinates $t, x, y$
\begin{equation}
 \Phi = \mu\,  {\rm tanh} (\lambda\mu z) \, ,
\label{kink}
\end{equation}
which is nothing but a domain wall separating two domains with the values of 
$\Phi = \mp \mu$. Actually, the domain-wall solution (\ref{kink})
satisfies the first-order differential equation
\beq
\partial_z \Phi + F = 0\, , \,\,\,
\mbox{where}\,\,\, F= -\lambda (\mu^2 -\Phi^2)\,  
.
\label{firstorder}
\eeq

The transverse size of the wall is of order $ (\mu\lambda)^{-1}$. A 
typical value of the mass of the non-zero modes is $ (\mu\lambda)$. We will 
assume that
$$
M\equiv  (\mu\lambda)
$$ 
is very large, while $\lambda^2\ll 1$, so that 
$m\equiv  \lambda^2\cdot (\mu\lambda)$
sets a ``normal" scale of the masses (energies) accessible
to the domain-wall dwellers. It remains to be added
that the energy density $\varepsilon$ of the wall (per unit area) is
\beq
\varepsilon = \frac{8}{3}\lambda_0 \mu^3_0 = \frac{8}{3}\lambda \mu^3 \, , 
\label{enden}
\eeq
where the subscript 0 marks bare quantities.

\subsection{SUSY and the wall}

The Wess-Zumino model, considered as a four-dimensional model,
is perfectly supersymmetric.
The infinitesimal supersymmetry transformations have the standard
form
\begin{equation}
\delta\Phi = \bar{\epsilon}\psi_+,~~~~
\delta\psi_+ = - \not \!\partial \Phi\epsilon_- + F\epsilon_+,~~~~
\delta F = -\bar{\epsilon} \not\!\partial\psi_+\, ,
\label{trans}
\end{equation}
plus similar transformations for $\Phi^\dagger$, $\psi_-$ and $F^\dagger$.
Here  $\epsilon$
is a transformation parameter, four-component real spinor. In our
notation for any spinor 
$$
\psi_{\pm} = {1 \pm i\gamma_5 \over 2}\psi \, .
$$

When we consider a given domain wall, Eq. (\ref{kink}), postulating that this 
particular
domain wall is our ``vacuum", translational invariance
in the $z$ direction and 
supersymmetry are spontaneously broken. More exactly, 
in the model at hand
a part of 
supersymmetry is broken. Indeed, two out of four transformations
do not act on the domain-wall solution. If
\beq
\gamma^3 \epsilon = -\epsilon
\label{eigen-}
\eeq
then, as   follows from Eq. (\ref{firstorder}),
$\delta\psi = 0$. The fact that
the given background preserves a part of supersymmetry has far-going
consequences. The energy of the solution is not renormalized
in loops, in much  the same way as  happens with
instantons \cite{NSVZ} (where also 1/2 of the supersymmetry is preserved).
This is why the first expression in Eq. (\ref{enden}) is exact;
the renormalizability of the four-dimensional theory then tells us
that $\mu^2 = \mu_0^2 Z$ and $\lambda = \lambda_0 Z^{-3/2}$,
with one and the same $Z$ factor, which leads to the second
expression  in Eq. (\ref{enden}), in terms of the renormalized quantities. 
Actually, 
this line of 
reasoning presents one of the possible proofs of the fact
that only the kinetic term is renormalized in the Wess-Zumino model,
cf. Ref. \cite{SV}. In other words, although the vacuum energy density
inside the domain wall is non-vanishing, the quantum loops do not change it.
 
Two other supersymmetry transformations, with
\beq
\gamma^3 \epsilon = \epsilon \, ,
\label{eigen+}
\eeq
act on the domain wall non-trivially. This means that
the corresponding components of the supercurrent are coupled to
the Goldstone fermion and effectively must be considered as broken.

This phenomenon -- breaking of a  half of  supersymmetry -- was  
observed
long ago in the  $(1 + 1)$-dimensional case \cite{wittenolive}, see also 
\cite{Bag,Mir,Ian}.
Our task is to analyse the impact of this breaking on the
$(2 + 1)$-dimensional Universe inside the wall. Correspondingly, we will
focus on 
the
low-energy world that originates from the massless modes trapped on the
membrane. There is a single normalizable real-scalar massless mode on the 
wall
$$
\phi (t,x,y,z)= a(t,x,y)\, \phi_0 (z)\, ,
$$
\beq
\phi_0 (z) = \sqrt{\frac{3\lambda\mu}{4}}\,
[{\rm cosh}(\lambda\mu z)]^{-2}\,  ;
\label{boszm}
\end{equation}
(it is assumed for definiteness that the parameters $\lambda$ and $\mu^2$
in the superpotential are real). 
This mode   corresponds to a small transverse shift of the membrane as a 
whole; 
in fact,  $a(t,x,y)$ is a Goldstone boson of the spontaneously broken 
translational symmetry.
The broken generator is $P_z$. Thus, the 
masslessness of one real boson is due to the Goldstone theorem and will be 
maintained not only in the tree approximation, but with all quantum 
corrections included. Moreover, 
 there is one ``chiral" massless fermion mode localized on the
membrane. This mode is obtained by applying to the wall solution
those generators of supersymmetry that act on it non-trivially (i.e.
the broken 
generators),
\begin{equation}
\psi(t,x,y,z) = 
\eta(t,x,y) \phi_0 (z) =\eta(t,x,y) {\rm e}^{- 2\lambda\int_0^z\Phi(z') dz'}\,
,
\label{fermzm}
\end{equation}
where  $\eta$ is a real spinor subject to the constraint 
\footnote{Sometimes we will write  $\gamma_z$ instead of  $\gamma^3$.}
$\gamma_z\eta = \eta$ 
(the term ``chiral" above is used just as a shorthand
for the real spinor subject to this constraint).
This fermion $\eta(t,x,y)$ is a 
Goldstone fermion of the broken half of  supersymmetry (goldstino). The 
constraint $\gamma_z\eta = \eta$ makes this spinor field two-component.
Indeed, with our choice of the $\gamma$ matrices, out of four components
in $\psi$ only two upper components survive in $\eta$.

The masslessness of $\eta$   is due to the Goldstone theorem as well. However, 
in contrast
with the scalar zero mode $\phi$, there is another circumstance
that keeps the fermion
massless, independently of its goldstino nature. 
An index theorem \cite{IT} tells us that the Dirac 
equation in the background of the static scalar field $\phi$
\begin{equation}
\left ( \not\!\partial - 2\lambda \phi \right)\psi = 0
\label{eqfzm}
\end{equation}
has exactly one normalizable localized zero-mode solution, whenever
the boundary conditions of the background field have the opposite signs
$\phi(+\infty) = - \phi(-\infty)$. This circumstance will be exploited below.

All other modes of the fields $\Phi$ and $\psi$ are massive, with mass
of order $M$. Now, we decompose the fields $\Phi$ and $\psi$ in modes,
and integrate out the heavy modes in order to get the low-energy effective 
Lagrangian for the fields $a$ and $\eta$. The tree-level Lagrangian is 
obtained by   substituting the
zero ones in the original Lagrangian (\ref{LWZM}), and integrating over $z$,
\beq
{\cal L}_{3} =
\frac{1}{2}\partial^\mu a \partial_\mu a
+ \frac{1}{2}\bar\eta \gamma^\mu\partial_\mu
\eta   \, ,
\label{L3}
\eeq
where $\mu$ runs over   0, 1, 2, the three-dimensional $\gamma$ matrices 
are defined as $\gamma^0 = i\sigma_2$, $\gamma^1 = -\sigma_3$,
$\gamma^2 = \sigma_1$.
Note that  the Lagrangian (\ref{L3}) 
describes a balanced number of boson and fermion degrees of
freedom -- one real boson field and one real (Majorana) two-component 
spinor. This pair forms an irreducible representation of supersymmetry
in three dimensions.  

The original $N=1$ supersymmetry, being viewed from three
dimensions, is $(N=2)_{d=3}$. The mechanism under consideration breaks it
spontaneously to $(N=1)_{d=3}$. Close problems with this
property were considered previously \cite{wittenolive,Bag,Mir,Ian}.
Related  suggestions as to how one may
spontaneously break $N=2$ down to $N=1$ are discussed
in Ref. \cite{Ant}. 

The ``minimal" model discussed above  is
our starting point. Below (in Sect. 2.3 and the subsequent sections) we suggest 
extensions where SUSY is
completely broken in the low-energy world. We consider the minimal
model mainly for the purpose of deriving a general criterion that would
allow one to build models with  completely broken SUSY
(see Sect. 2.2).

 Integrating
out the non-zero modes may give rise to counter-terms proportional to
$g^2$,
say, quadratic in $a$. The net effect of these counter-terms is
to ensure that the physical mass of the $a$ field stays at zero even
after inclusion of the interactions given in Eq. (\ref{L3}). 
The same is valid for 
counter-terms with the fermion field.

\subsection{1/2 of SUSY in $d=3$ from SUSY in $d=4$}

Since the emerging three-dimensional theory is supersymmetric,
the standard SUSY algebra 
\beq
\{ \bar Q_\alpha^{(3)} Q_\beta^{(3)}\} = 2 (\not\! p)_{\alpha\beta}
\, , \,\,\,  
\alpha 
,\beta = 1,2 
\label{3alg}
\eeq
must take place.  Here $Q$ is a real two-component
supercharge and the superscript 3 indicates that the quantities refer to the 
three-dimensional effective theory; $p$ denotes the
momentum operator in the  three-dimensional  theory.
It is instructive to trace how this algebra 
could
appear  from the superalgebra of the original theory 
where the supercharge $Q$ was a four-component spinor.

The key point is that the four-dimensional theory we consider
has superalgebra with the central extension,
\beq
\{ \bar Q_\alpha^{(4)} Q_\beta^{(4)}\} = 2 (\not\! P^{(4)})_{\alpha\beta}
+ 2 (\gamma^5\sigma^{\mu\nu} |J^{\mu\nu}|)_{\alpha\beta} 
\, , \,\,\,  \alpha ,\beta = 1,2,3,4 
\label{4alg}
\eeq
where $\sigma^{\mu\nu} =(1/2)[\gamma_\mu ,\gamma_\nu ]$ and 
$J^{\mu\nu}$
is a trivially conserved (topological) charge,
\beq
J^{\mu\nu} =\int d^3 x \varepsilon^{0\mu\nu\rho} \partial_\rho  W(\Phi )\,  ;
\label{topcur}
\eeq
$P$ is the
momentum operator in the  four-dimensional  theory.
The conservation of the current $\varepsilon^{\sigma\mu\nu\rho} 
\partial_\rho W(\Phi) $ is obvious. Usually it is believed that $N=1$ 
superalgebra (and the one we deal with in the four-dimensional theory
(\ref{LWZM}) is $N=1$ from the point of view of four dimensions)
cannot have central extensions. The standard proof includes two ingredients:
(i) the scalar conserved central charge can not appear in the
$N=1$ superalgebra; (ii) the Coleman-Mandula theorem
\cite{CM} forbids the existence of the conserved quantities
with the (vector, tensor, ...) Lorentz indices, other than four-momentum,
in dynamically non-trivial theories. Our topological charge $J^{\mu\nu}$
is the antisymmetric tensor. The Coleman-Mandula theorem is avoided
due to the fact that $J^{\mu\nu}$ is non-vanishing only if the translational  
symmetry is spontaneously broken, as is the case with the domain wall.
In  the absence of the breaking  of translational invariance, $J^{\mu\nu}=0$,
in full accord with Ref. \cite{CM}.

Now, we start from Eq. (\ref{4alg}) and reduce it to our effective low-energy 
three-dimensional theory. Let us assume that we are in the rest frame of the
domain wall. For any state from the Hilbert space of the low-energy 
three-dimensional theory 
$$P_0= M_{\rm wall} + p_0 \, , \,\,\,\,
P_{1,2} = p_{1,2}\, ,\,\,\,  P_3=0
\, ,
$$
 where  $M_{\rm wall} =
\varepsilon V_2$. Moreover, the only surviving central charge
is $J^{12}$, its value is $M_{\rm wall}/2$, and 
 it cancels the same term appearing in the transition from the
four-dimensional momentum to the effective three-dimensional
one for  appropriately chosen values
of the spinor indices $\alpha , \beta$ (corresponding
to the conserved supercharges). In this way Eq. (\ref{4alg}) is reduced 
to Eq. (\ref{3alg}).

Thus, we arrive at a useful working criterion: if a part of a translational 
invariance is broken by a domain wall or a similar defect, {\em and}
the theory we start from has {\em no} (non-vanishing) central charges,
full supersymmetry has to be spontaneously broken. It is very easy to modify 
the minimal model in such a way that $J^{\mu\nu}=0$, although
the domain wall still exists. This is done in the next section.

\subsection{Non-minimal wall}

The model considered above is a minimal SUSY model allowing for the
wall solution. The fact that the fermion zero modes on the wall can appear
due to the index theorem, not necessarily related to goldstinos,
gives us a hint that it may be possible, by an appropriate extension of the 
model, 
to have  fermionic zero modes  not accompanied
by the scalar zero modes. 
In such a non-minimal extension, SUSY must be -- and actually is --
fully broken in the effective low-energy theory. The analysis
of Sect. 2.3 tells us that to completely break supersymmetry
one must ensure the vanishing of the central charge
$J^{\mu\nu}$ on the wall solution. 

Let us consider the simplest example. 
Introduce a number of ``quark"
superfields $Q_A$ ($A = 1,2,..., n$) and an additional superfield $X$.
Consider a superpotential
\begin{equation}
W = X\lambda(\Phi^2 - \mu^2) + {h_A\over 2}\Phi Q_A^2\, .
\label{WXQ}
\end{equation}
It is invariant under the discrete symmetry $\Phi \rightarrow -\Phi$,
$Q_A \rightarrow iQ_A$. The vacuum of the four-dimensional theory is at
\begin{equation}
\Phi = \pm \mu,~~~X=Q_A=0\, .
\label{extvac}
\end{equation}
If $X$ and $Q_A$ vanish the non-linear equation on $\Phi$
is exactly the same as in the minimal model discussed above, and
it has the same wall solution (\ref{kink}). However, now $J^{\mu\nu} =0$,
and SUSY is completely (spontaneously) broken.

First, let us show that  there are no combinations
of the super-transformations that act on the wall solution trivially
(i.e. annihilate it). Fermionic components of $X$ and $\Phi$
superfields in the given background field are
\beq
\delta\psi = - \partial_z \phi\gamma_z\epsilon_-,~~~~
\delta\psi_X = F_X\epsilon_+\, ; 
\label{bshift}
\eeq
(recall that $\partial_z X = \ F_{\Phi} = 0$ for the wall solution
while $\partial_z \phi\neq 0$ and $F_X\neq 0$).
Thus, for any $\epsilon \neq 0$ the fermion is created out of the wall-dweller 
(three-dimensional)  ``vacuum",
and supersymmetry is fully broken. Correspondingly,
there occurs a {\em complex} two-component goldstino --
a mixture of $\psi_X$ and $\psi_\Phi$, see Eq. (\ref{bshift}).
In  view of this fact, it
is not surprising that the numbers of the fermionic and bosonic
zero modes are different on the wall in the model at hand.

Indeed, 
 the Fermi components of $Q$,
$\psi_{Q_A}$,
satisfy the Dirac equation, similar to (\ref{eqfzm}), which by the index 
theorem 
\cite{IT} has a
single zero-mode solution for each $\psi_{Q_A}$
\footnote{The very same observation allows
one to simulate chiral massless fermions on four-dimensional lattices starting 
from a lattice theory of massive
interacting fermions in five dimensions, provided that the fermion mass
has a step function shape in the extra dimension. This is
the so-called Kaplan chiral lattice fermion \cite{Kaplan}.}. The  bosonic 
counterpart of this equation is
\begin{equation}
\partial_{\mu}\partial^{\mu}Q_A + h_A^2|\Phi|^2Q_A = 0 \,  .
\label{nbzm}
\end{equation}
Subsequently, the equation for the $z$-dependent part is a Schr\"odinger
equation with a semipositive-definite potential and has no
zero frequency bound-state solution. Thus, the corresponding scalar zero 
modes  of the ``squark" fields are
absent on the membrane. 
The only boson zero mode is the Goldstone mode of the $\Phi$ field itself,
associated with the spontaneous breaking of $P_z$. It is the same mode we 
dealt with in Sect. 2.1. A disbalance between the
number of massless bosons and fermions is obvious.
The three-dimensional observer living on the membrane
would not observe any Fermi-Bose degeneracy in the spectrum.

Since the fermion zero modes in the case at hand are not goldstinos,
the corresponding particles may acquire masses through loops. These masses,
however, will be small, of order $m$, not $M$.

\section{Compactification on the Global Cosmic String}

The above example 
is the  existence proof of  complete SUSY breaking through 
dynamical compactification. Below we will discuss several other theories
where this phenomenon takes place.

Here we extend the strategy   to the case of the Abelian
global $U(1)$ symmetry, with a $(1 + 1)$-dimensional cosmic string solution.
(Such global strings were discussed previously, e.g. in Ref.
\cite{Vil}.)
Two dimensions are dynamically compactified, say $x$ and $y$
(we assume that the string is in the $z$ direction). The global 
$U(1)$ symmetry is obtained by introducing  an additional superfield 
$\bar\Phi$ 
in Eq. (\ref{WXQ}). Then the 
superpotential takes the form
\begin{equation}
W = X\lambda(\Phi\bar{\Phi} - \mu^2) + {h_A\over 2}\Phi Q_A^2\, .
\label{WXQbar}
\end{equation} 
The  $U(1)$ symmetry
$$
\Phi \rightarrow \Phi{\rm e}^{i\alpha},~~~~~
\bar{\Phi} \rightarrow \bar{\Phi}{\rm e}^{-i\alpha},~~~~~~
Q_A \rightarrow  Q_A^{-i{\alpha / 2}}
$$
is explicit. The vacuum expectation values of the fields $\Phi$ and
$\bar\Phi$ break this $U(1)$ spontaneously. Note that the 
vacuum expectation values of $X$ and $Q_A$ do not develop.

This theory admits a topologically non-trivial global vortex-line solution;
 in the cylindrical coordinates $(\rho, \theta)$  this  has the form
\cite{Vil}
\beq
\Phi = {\bar{\Phi}^*} = f(\rho){\rm e}^{in\theta} \, ,
\label{vortex}
\eeq
where $\rho$ is the distance from the $z$ axis in the $\{ x,y\}$ plane
and $\theta$ is the azimuthal angle with respect 
to the $z$ axis. In what follows we assume  that $n=1$,
since the vortex is stable in this theory only for the minimal winding.

Equation (\ref{vortex}) is a $(1 + 1)$-dimensional cosmic string
oriented along 
the $z$
axis. The profile function $f(\rho)$ is a smooth function such that $f(\rho ) 
\propto \rho $ at small $\rho$;
moreover,  $f(\rho)$ approaches
$\mu$  (we assume $\mu$ to be real) as 
\beq
f(\rho) -\mu \sim \frac{1}{\rho^2}\,\,\,\mbox{at}\,\,\, \rho\ra\infty\, .
\label{asy}
\eeq
Equation (\ref{asy}), as well as the fact that the absolute values of
the fields $\Phi$ and $\bar\Phi$ must be the same on the vortex solution,
is readily obtained from the classical equations of motion. At large $\rho$
they reduce to
$$
\frac{1}{\rho^2}\bar\Phi +\lambda^2 (\bar\Phi\Phi -\mu^2)\Phi^* = 0\, , \,\,\,
\frac{1}{\rho^2}\Phi +\lambda^2 (\bar\Phi\Phi -\mu^2)\bar\Phi^* = 0\, .
$$
 As a result, the 
$F$ terms vanish as $1/\rho^2$  away from the defect; the corresponding
energy vanishes as $1/\rho^4$.  
The gradient energy density due to the $D$ terms is 
\beq
|\partial_{\mu}\Phi|^2 \rightarrow
\left (\mu /{\rho} \right )^2 \, ,\,\,\, \mbox{at}\,\,\, 
\rho \rightarrow \infty \, .
\eeq
It vanishes  more slowly and is dominant.

Once again, the fermion transformations in the string background
indicate that supersymmetry is completely broken by this solution,
$$
\delta (\psi_{\Phi})_{ +} =
- {\rm e}^{i\theta}\left(
\gamma_{\rho}\partial_{\rho} f(\rho) + {if(\rho) \over \rho}
\gamma_{\theta}\right) \epsilon_-\, ,\,\,\, 
\delta (\psi_{\bar{\Phi}})_{ +} =
- {\rm e}^{-i\theta}\left(
\gamma_{\rho}\partial_{\rho} f(\rho) - {if(\rho) \over \rho}
\gamma_{\theta}\right) \epsilon_-\, ,
$$
\beq
\delta\psi_X = F_X\epsilon_+ \,  ,
\label{sshift}
\eeq
where $\gamma_{\rho} = \gamma_1 {\rm cos}\theta + \gamma_2{\rm 
sin}\theta$
and $\gamma_{\theta} = \gamma_1 {\rm sin}\theta - \gamma_2{\rm 
cos}\theta$.
The right-hand side of these equations cannot vanish for any choice of the
transformation parameter $\epsilon$.
As in the case of the non-minimal wall, the supersymmetry breaking
on the string manifests itself in a disbalance between  the Fermi and Bose
zero modes. By the index theorem,  a non-trivial winding of the Higgs-field
phase results in a (single) massless fermionic mode for each fermion field
which in the vacuum gets a mass from the Yukawa coupling to the Higgs
in question \cite{wjr}.
These modes satisfy the following $z$0independent Dirac equation,
\begin{equation}
\gamma_1({\rm cos}\theta + \gamma_1\gamma_2{\rm sin}\theta)
{\partial\psi_+ \over \partial\rho} = - 2 h_A f(\rho) 
{\rm e}^{i\theta}\psi_-
 \label{dir}
\end{equation}
(where $\psi_+$ runs over all the fermions coupled to the defect and
$h_A$ is the  corresponding Yukawa interaction constant).
The above equation is satisfied by
\begin{equation}
\psi =\beta(z - t) \eta{\rm e}^{-2h_A \int_0^{\rho} f(\rho')d\rho}\, ,
\label{rmover}
\end{equation}
where $\eta$ is a constant spinor such that $\eta = 
\gamma_1\gamma_2\gamma_5\eta$ and $\eta 
=\gamma_1\eta$. The solution  (\ref{rmover}) describes a single chiral
(right-moving) zero mode propagating along the string.

The set of the bosonic zero modes trapped on the vortex includes only  two 
Goldstone modes
corresponding to the transverse displacements of the string in the
$\{x,y\}$ plane.

\section{Gauge String Example}

It is not difficult to construct an
analogous example of dimensional reduction from
$3 + 1$ to $1 + 1$ in the case when the $U(1)$ symmetry is gauged. The 
reduced 
two-dimensional world is the inside of  an infinite {\em local} cosmic string.
Consider an $N=1$ globally supersymmetric theory with the $U(1)$ gauge 
symmetry, a version of supersymmetric electrodynamics (SQED).
We introduce a single chiral superfield $\Phi$ with the  charge $-2$ and
eight ``quark" superfields $Q_A$ with the unit charges. The number of the 
``quark"  fields is 
dictated merely by the requirement of the anomaly cancellation 
in SQED and is in no way
essential for our analysis \footnote{In this example  the trace of the 
charges is nonzero, ${\rm Tr} \, q = 6$, implying a mixed $U(1)$-gravitational 
anomaly were the model embedded 
in the 
supergravity framework. Again, the condition ${\rm Tr}\, q \neq 0$ is not
essential for our purposes,  and by adding extra fields we could,
in principle, arrange the vanishing of ${\rm Tr}\,  q = 0$. Our conclusions
are independent of the ${\rm Tr}\, q$ condition.}. The gauge section of
the model at hand has a
global $SU(8)$ symmetry. Using this symmetry one
can always reduce the most generic superpotential to a diagonal form, by an
appropriate 
 field redefinition,
\begin{equation}
W = \sum_A \, \frac{g_A}{2} \Phi Q^2_A\, .
\end{equation}
Then  the scalar potential takes the form
\begin{equation}
V = \frac{g^2}{2}\left ( -2 |\Phi|^2 + \xi^2 + |Q_A|^2 \right )^2 +
g_A^2|\Phi|^2|Q_A|^2 + \frac{1}{4} |g_AQ^2_A|^2\, .
\end{equation}
Here $g$ is the  gauge coupling constant and $\xi^2$ is a Fayet-Iliopoulos
$D$ term which we take to be positive \footnote{In the present example,
the Fayet-Iliopoulos $D$ term,
even if it is zero at the tree level,  will be generated at one
loop \cite{oneloop},
 proportional to
${\rm Tr} \, q$. Note, that beyond one loop this term is not renormalized,
see \cite{oneloop,SV}.}. Minimization of the scalar potential yields
\begin{equation}
|\Phi|=\frac{1}{\sqrt{2}}\xi,~~~|Q_a|^2=0\, ,
\end{equation}
and the $U(1)$ gauge symmetry is spontaneously broken by the vacuum 
expectation value of $\Phi$.
It is well known \cite{string} that such a system admits a
topologically stable solution, the  
 Abrikosov vortex line, or the Nielsen-Olesen string,
which in  cylindrical
coordinates has a form similar to (\ref{vortex})
(again we assume the string to be oriented along the $z$ axis),
\begin{equation}
\Phi = f(\rho){\rm e}^{in\theta}\, ,
\end{equation}
where $f(\rho) \sim \rho$ at $\rho \rightarrow 0$ and $f(\rho)\rightarrow 
\mu$ at $\rho
\rightarrow \infty$.  The winding in phase is accompanied by
the corresponding solution for the gauge field $A_\mu$. The difference from 
the global
case is that, away from the core, the gradient energy is now compensated 
by the gauge vector
field $A_{\mu}$ which assumes a pure gauge form at  infinity, 
$A_{\theta} \rightarrow -n/\rho$. Also, the parameter $n$ need not be equal 
to unity; higher windings are stable as well. Consequently, a magnetic flux 
becomes
trapped in the core of the defect.

Let us consider the zero modes on the Abrikosov--Nielsen--Olesen
vortex. The modes from 
the supermultiplets $Q_A$ are of special interest. Their fermionic components
satisfy one and the same Dirac equation in the string background
\begin{equation}
 \not \! \! D\psi^A_-= 2 g_A \Phi \psi^A_+\, ,
\end{equation}
which by the index theorem has $n$ normalizable zero-energy solutions 
trapped
on the string \cite{wjr}. The corresponding scalars $Q_A$  have no zero modes, 
since their
mass-squares are positive-definite. There are, of course, the bosonic zero 
modes associated with the displacement of the string as a whole; they involve
both $\Phi$ and the vector gauge field. The number of the fermion zero 
modes, however, grows with the number of the ``quark" superfields and with
the winding number of the string, and cannot be balanced. SUSY is completely 
broken as in the global case. 

\section{Solutions with Purely Gradient Energy}

In this section we consider the dynamical  compactification via classical
solutions, which -- although stable under  finite deformations -- are {\em not}
topological defects in the usual sense. The key differences are:
(i) the expectation value of the Higgs fields never leaves the  vacuum 
manifold,
and (ii) its absolute value is not asymptotically constant at  infinity.
Such solutions are only possible thanks to the existence of  non-compact 
flat
directions in the supersymmetric theories  and have no analogs in
non-supersymmetric theories. As far as we know, this type of the vacuum 
defects
is novel; they have not been considered previously. Their dynamical impact,
in some aspects, is similar to that generated by the non-minimal wall
(Sect. 2.3), but it is different in other aspects. As a matter
of fact, one can 
obtain the peculiar solutions
at hand from a conventional wall, by a limiting procedure (see the end of this 
section).

Since the Higgs field is on the vacuum manifold
everywhere, the configuration at hand can only contribute  to the 
gradient energy,
and the Bogomol'nyi equations cannot be defined. As a result, the 
configuration breaks supersymmetry and (partially) the translational 
invariance.

Let us discuss  simplest solutions of this type.
Consider an $N=1$ supersymmetric model  with two chiral
superfields $\Phi$ and $Q$, and the superpotential of the form
\begin{equation}
W = {h \over 2}\Phi Q^2\, .
\label{dsn}
\end{equation}
The vacuum is at $Q = 0$ and $\Phi$ undetermined. So, $\Phi$ can
assume any value (as long as $Q$ vanishes) without contributing to the
potential energy. In other words,  we have a vacuum valley along the $\Phi$ 
direction. 

Now, a new stable configuration breaking translational invariance
(analog of the wall) is as follows.
We may think of the configuration in which $\Phi$
increases with coordinates linearly when we travel from minus to
plus infinity.
Such configurations, indeed, do  satisfy the equations of motion.
Depending on the behaviour of the phase, we can distinguish two cases.

\vspace{0.2cm}

{\it Constant Phase Configuration}

\vspace{0.1cm}

 This can be written simply as 
\begin{equation}
\Phi = \alpha z\,  ,
\label{Zgorka}
\end{equation}
where for definiteness $\alpha$ is assumed to be a real number.
The translational invariance in the $\{x,y\}$ plane is obviously preserved,
while  it is broken in the $z$ direction. 

It can easily be checked  that the solution (\ref{Zgorka}) is stable under any 
finite
deformations. The vacuum energy density is given by a constant gradient 
energy
$\varepsilon = \alpha^2$, which is nothing but the expectation value  of the 
$D$ term
$\int d^4\Theta \Phi^*\Phi$ ($\Theta$ is a superspace coordinate).
As was mentioned, there is no contribution 
to $\varepsilon$ from the $F$ terms.

Our surrogate ``vacuum" (\ref{Zgorka}) breaks supersymmetry completely. 
Indeed, it is not difficult to  see that for any non-zero $\epsilon$ a
fermion is created out of the ``vacuum". This is because $\gamma_z$ has no
zero eigenvalues. Another argument leading to the same conclusion
is the vanishing of the central charge in the model at hand. 
The corresponding goldstino is the  fermion component of $\Phi$. Both the 
real 
and imaginary parts
of the massless scalar $\Phi$ are Goldstone bosons:
one is due to the spontaneous breaking of the 
translational invariance (in the $z$ direction), and another 
due to the breaking of the global
$U(1)$ symmetry,
\begin{equation}
\Phi \rightarrow {\rm e}^{i\omega} \Phi;~~~
Q \rightarrow {\rm e}^{-i\omega /2} Q \, . 
\label{u1}
\end{equation}
The Goldstone particles and the goldstinos  are not confined to the two-
dimensional 
surface $z = 0$;
the corresponding modes are not localized in the $z$ direction, and there
is no mass gap for these particles.

Does the configuration (\ref{Zgorka}) compactify the four-dimensional world?

Yes and no. In a sense, it compactifies, if we assume that the 
three-dimensional observers 
are made from the ``quark" quanta, from the superfield $Q$. These quanta
are localized near the two-dimensional surface $z = 0$.
Consider the massless particles localized on the membrane. First, there are no
localized massless scalars on the wall. 
The scalar component of
$Q$ obeys the  oscillatory equation
\begin{equation}
\left (-\partial_z^2 + h^2 \alpha^2 z^2\right ) Q = 0\, ,
\end{equation}
with no localized {\em zero-mode} solution. Actually, all solutions are 
localized,
and have mass $\sim {h}\alpha$.

In contrast,
trapped fermionic zero modes do exist. This is the  fermionic component of
$Q$. The corresponding normalizable solution has the form
\begin{equation}
\psi = \eta(t,x,y){\rm e}^{- \alpha h z^2}\, .
\end{equation}
As before, $\eta$ is the  $ + 1$ eigenvalue of $\gamma_z$. 
The three-dimensional observers are made from these quanta.
These quanta interact with the delocalized modes from $\Phi$, the
Goldstone particles.  Thus, the three-dimensional 
observer could, in principle, see the leakage of the three-momentum
into the extra dimension; unlike
the standard wall considered in Sect. 2.3,   
this    
                             leakage
is not a threshold effect since there is no mass gap for the modes of the
$\Phi$ field. 
Interaction of the $\eta$ particles with the $\Phi$ quanta can be made 
arbitrarily weak, however,  provided that $h\ra 0$ and $\alpha$ is scaled 
appropriately.

Needless to say, we could have introduced several ``quark"
superfields, each with its own coupling to $\Phi$, interacting with each other,
$$
W = {h_A \over 2}\Phi Q^2_A +g_{ABC}Q_AQ_BQ_C\, .
$$
If $g\gg h$ the dominant interaction of the $\eta$ particles will appear
after integrating out the non-zero modes and will be of the
form $g^2 (h\alpha)^{-1} \bar\eta\eta\bar\eta\eta$.

The solution (\ref{Zgorka}) can be viewed as a limiting case
of the conventional domain wall in the non-minimal model (\ref{WXQ}),
with $\lambda \ra 0$ and $\lambda\mu^2$ fixed (and equal to $\alpha$),
see Eq. (\ref{kink}).
If $\lambda\neq 0$ the vacuum degeneracy is lifted, there is no
flat direction, and the standard wall interpolates between
$-\mu$ and $+\mu$ vacua at $-$ and $+$ infinities in the
$z$ direction. When $\lambda \ra 0$ the vacuum degeneracy is restored, 
while $\mu\ra \infty$. The inside of the wall now spans the whole space. 
Taking this limit is useful in order 
to understand the  nature of the solution; there is a subtle point,  however,
which one should keep in mind  to obtain the theory (\ref{dsn}) from 
(\ref{WXQ}): in this limit (\ref{WXQ})
gives a sterile superfield $X$,  with  a linear term in the superpotential,
$X\alpha$,  which, although irrelevant,   also contributes to the vacuum 
energy.

\vspace{0.2cm}

{\it Winding phase}

\vspace{0.1cm}

Since the non-zero VEV of $\Phi$ spontaneously
breaks the $U(1)$ global symmetry (\ref{u1}), there is a cylindrically
symmetric solution with a winding phase,
\begin{equation}
\Phi=\alpha\rho{\rm e}^{i\theta}\, . 
\label{wind}
\end{equation}
This solution reminds the  global cosmic string, except for  the fact that
the modulus of the Higgs field
never assumes a constant value.
Due to the winding number, there is a normalizable zero mode of $\psi_Q$
trapped at the origin, 
\beq
\psi_Q = \beta(z - t)\eta{\rm e}^{- h\rho^2}\, ,
\eeq
which simply represents a solution of the Dirac equation 
(\ref{dir})
with $h_Af(\rho) = h\rho$.
Other zero modes are: (i) the ({\it non-localized}) massless
complex scalar field
$\Phi$ with two real components being the Goldstone bosons of the broken
translational invariance and the $U(1)$ symmetry; (ii) the ({\it non-localized})
massless four-component Majorana fermion $\psi$, the goldstino. 

\section{SUSY Breaking through  Global 
Winding}

 The mechanism to be  considered in this section is somewhat different
from those discussed above, since in this
case the compactification
is not dynamical. Rather, as in the conventional Kaluza-Klein scenarios,
we choose  the space-time manifold of the
given topology, say, with one compact dimension, from the very beginning. 
However, once this is done, the supersymmetry breaking in
the effective low-energy theory does occur {\it dynamically}, as a result of the 
non-trivial homotopy of the space-time and the vacuum manifold of the
spontaneously broken internal symmetry.
The simplest example we can think of 
in this connection 
is the one with
spontaneously broken global symmetry (\ref{WXQbar}). Consider the model of 
Sect. 3; we no longer need  the ``quark" fields $Q$, however, and we
exclude them for simplicity. Assume that one of the coordinates,
say $z$, is compact with the  topology of a circle $S_1$ of  radius $R$,
so that the four-dimensional  space-time manifold is a cylinder.
Now, since the 
first homotopy
groups of both the vacuum (i.e. $U(1)$) and the space-time manifolds
are non-trivial, there is a topologically stable $(t,x,y)$-independent
winding configuration
\begin{equation}
\Phi = \bar{\Phi}^* =\sqrt{\frac{\mu^2 R^2\lambda^2- 
1}{R^2\lambda^2}}
\, {\rm e}^{i{z \over R}}\, .
\label{zmeika}
\end{equation}
Equation (\ref{zmeika}) is obviously a solution of the classical
equations of motion, subject to the constraint of the unit global winding.

Defining
\beq
\Phi = \phi(x,y,t){\rm e}^{i{z \over R}}\,\,\, \mbox{and}
~~~\bar{\Phi} = \bar{\phi}(x,y,t){\rm e}^{-i{z \over R}}
\eeq
and integrating over the compact coordinate, we arrive at an effective
$(2 + 1)$-dimensional theory in which the $\phi$ and $\bar\phi$ fields get 
the
following contribution to the masses from the gradient energy density in the 
$z$ direction
\beq
m_{\phi} = m_{\bar{\phi}} = \frac{1}{ R}\, .
\label{soft}
\eeq
This mass {\it per se} does not break supersymmetry. SUSY is broken, 
however,   
if the vacuum expectation values of $\phi, \bar{\phi}$ are non-zero (as  
happens
 in our model), in which case they induce a non-vanishing expectation
value of the auxiliary component of the  superfield $X$,
\beq
F_X= { 1 \over \lambda R^2}\, .
\eeq
The latter breaks supersymmetry spontaneously. Another quantity that 
contributes
to the spontaneous supersymmetry breaking is the expectation value of the
$D$ term:
\beq
\left\langle \int d^4\Theta \Phi^*\Phi \right\rangle =
\frac{\mu^2 R^2\lambda^2- 
1}{R^4\lambda^2}\, ,
\eeq
 setting the scale of SUSY breaking in the low-dimensional world.
Note that in the limit $R \rightarrow 0$ ($\mu$ and $ \lambda $ fixed)
supersymmetry is restored. This is because the winding
configuration in Eq. (\ref{zmeika}) becomes unstable for $R \ll 1/\mu$: the 
gradient energy 
becomes
so strong that is can pull the Higgs fields over the potential barrier
and unwind the solution.

Let us briefly discuss the Goldstone modes.
First, in the model at hand the
solution (\ref{zmeika})
breaks the invariance under displacements $z\ra z + r$. It also breaks
 the invariance under the global $U(1)$ rotations.
This is one and the same 
breaking, however, since the two transformations are entangled.
Correspondingly, the Goldstone mode is a single superposition (the difference)
of the $\Phi$ and $\bar{\Phi}$ phases.

What about goldstino? The criterion of the vanishing central charge
tells us that SUSY is completely broken. Correspondingly, the
goldstino field is a complex two-component field, a mixture
of $\psi_\Phi$, $\psi_{\bar\Phi}$ and $\psi_X$. It is instructive
to find the goldstino combination explicitly. To this end it is convenient
to decompose the spinors in terms of the $\pm 1$ eigenspinors of 
$\gamma_z$.
Then  the corresponding components of the goldstino (up to a normalization
factor) are given by the following superpositions of the initial fermions:
\beq
\psi_{\rm goldstino}^{\pm} = {1 \over \lambda R^2}\psi_X^{\pm} \pm
{i \over R}\sqrt{\frac{\lambda^2\mu^2 R^2- 1}{\lambda^2R^2}}
\left ( \psi_{\Phi}^{\pm}{\rm e}^{-i{z \over R}}  -
\psi_{\bar{\Phi}}^{\pm}{\rm e}^{i{z \over R}} \right )\, .
\label{KKgold}
\eeq
These components are created out of the vacuum by the action of 
$\epsilon_{\pm}$
supersymmetry transformations, respectively.

The above mechanism is practically ready-made for exploitation in
phenomenological models. 

Superficially the  mechanism considered in this section might look similar to 
that discussed in Refs. 
\cite{ss,rohm} (the 
coordinate-dependent 
compactification). The similarity does not extend too far, however.
Indeed, within the coordinate-dependent 
compactification the boundary conditions on the boson and fermion
components are different,
much in the same way as in the high-temperature case \cite{HT}. In our 
approach 
the boundary conditions are the same (periodic) for bosons and fermions, and
SUSY breaking occurs because of the non-vanishing gradient energy carried 
by
the topologically stable configuration. Identical Fermi-Bose boundary
conditions follow from the fact that all fields wind by
the ({\it spontaneously broken}) $U(1)$ transformation commuting
with supersymmetry. Supersymmetry gets restored whenever the order 
parameter
that breaks $U(1)$ vanishes, or the winding configuration becomes unstable.
As shown above, the  instability happens when $R \rightarrow 0$.
A closer parallel may be found with the constant magnetic field mechanism
of Ref. \cite{magnet}. The basic distinction is that the latter deals with
the string theory compactification.

\section{Uses for Phenomenology}

There is  hope that at least some of the  mechanisms of  spontaneous SUSY 
breaking from the variety we suggest may prove to be relevant for 
phenomenology. These mechanisms are quite distinct from those
considered previously; in some  aspects they are less restrictive,
which opens many new opportunities. In particular, one can easily avoid
the standard supertrace relations. In this paper it seems premature
to submerge in excessive detail. We will briefly discuss   only
the global winding mechanism of Sect. 6, which is readily adjustable
for phenomenological purposes.
Let us show how one can exploit this mechanism in the
context of the models with the gauge-mediated SUSY breaking, rather
popular at present \cite{GM}.

The general modern approach to SUSY breaking is as follows.
SUSY breaking occurs in some alien sector.
Once originated, it can be transmitted to 
``our"  observable fields through some messenger interaction that will
not restore supersymmetry on its way. In our case, the simplest possibility is 
a 
direct coupling of some of the visible sector fields $Q_A$ to $\Phi$,
as in Eq. (\ref{WXQbar}). The non-zero Fermi-Bose splitting in $\Phi$ results
in  non-zero one-loop soft masses  of $Q_A$.

Perhaps a better candidate for the messenger is a gauge interaction,
as  happens in the  usual gauge-mediated scenarios \cite{GM}.
Such a scenario can be readily implemented within the  global winding 
scheme of 
supersymmetry
breaking suggested in Sect. 6.
All one needs is to introduce, additionally,  two 
pairs of the 
messenger superfields
$M$, $\bar M$ and $M'$, $\bar M'$ that transform under a gauge group 
$G$. Why we need two pairs, not one, will become clear shortly. The 
superpotential
can be written as
\beq
W = \lambda X \left ( \Phi \bar{\Phi} + \lambda_1M\bar M -\mu^2 \right )
+ \left ( M'\bar M\lambda' + \bar M' M \lambda''\right ) \mu
\eeq
where  $\lambda_1$, $\lambda'$ and $\lambda''$ are
parameters. It is easy to see that for arbitrary non-zero values of
the parameters  the above system can never restore  supersymmetry, which 
is
spontaneously broken for the non-trivial winding configurations, as discussed
in Sect. 6. As a matter of fact, integrating over the extra dimension, in the 
effective
 low-dimensional
theory we get the O'Raifeartaigh-type supersymmetry breaking. This analogy,
however, is not complete since the winding that results in  the
non-vanishing $D$ terms
plays a crucial role in this breaking. Assuming for definiteness that
\beq
  \lambda'\lambda''\mu^2R^2 > \lambda_1
\eeq
we check that  minimization gives
\beq
 M = \bar M = M'=\bar M' = 0\, .
\eeq
The only non-vanishing $F$ term is that of $X$.
(The second pair of fields, $M'$,
$\bar M'$, was introduced just for this purpose -- ensuring
that $F_X\neq 0$. Otherwise, the vacuum expectation value
of $M\bar M$ would adjust itself in such a way as to kill $F_X$.)
The non-vanishing $F_X$  splits the masses
of the Fermi-Bose components in $M, \bar M$. This splitting will be 
transmitted
to all $G$ nonsinglet states through the universal
two-loop diagram \cite{GM}. 

\section{Conclusions}

We find it  intriguing to think that two major questions in the modern theory 
-- the origin of our four space-time
dimensions and the origin of SUSY breaking -- are related, and that both 
phenomena can be viewed
as a product of one and the same mechanism. The idea of our Universe being a 
topologically (or 
non-topologically) stable defect in higher-dimensional space-time is
particularly attractive in this respect. It is remarkable that 
the dynamical compactification with an automatic spontaneous breaking of all 
supersymmetries is possible, 
at least at
the level of toy models presented in this paper.
Although in the
vacuum of the original higher-dimensional theory, SUSY 
is fully operational,   it is spontaneously
broken inside   the defect. For the 
low-energy 
observers 
made from the zero-modes quanta
trapped in the core of the defect, the four-dimensional Universe
will look perfectly non-supersymmetric.

An important difference of the mechanism we suggest, with respect to the 
usual 
Kaluza-Klein-type compactifications on
the compact manifolds, is the absence of the infinite ``tower" of massive
excitations. This is because  extra dimensions are not really compact
but rather hidden beyond the potential barrier. The compactification scale 
(the height of the
barrier) can be very low, say, $\sim 1$ TeV or so, which is accessible,
in principle, to the current generation of 
colliders.
In such a picture, above a certain threshold energy, one
must see in  collisions, simultaneously
with the production of superpartners,  a 
leakage of
 energy-momentum  to extra dimensions.

The cosmological history of such a ``defect"  Universe must be very different 
from the conventional scenarios too.
In particular, it can be formed as a result of a phase transition
via the Kibble mechanism \cite{Kib} in the Universe of a higher 
dimensionality.
This is absolutely transparent for the wall-like and vortex-like solutions. 
Thus, in our approach the
 dimensional reduction is a phenomenon  that may happen in
time. Once produced as a network of, say, the domain walls or strings,
the low-dimensional Universe will evolve in a way very different
from the conventional big-bang scenario. In particular the surface tension
will play an important role, so that large clusters will try to straighten
out under this force \cite{ZKO}, approaching the state of asymptotically zero
curvature (without inflation).

In short, we think that the above possibilities, although quite speculative at 
the present stage, may prove to be fruitful. They  are worth studying in a 
more
realistic context.

\vspace{0.3cm}

{\bf Acknowledgments}: \hspace{0.2cm} The authors
would like to thank S. Ferrara, P. Hazenfratz, E. Kiritsis, I. Kogan, A. Kusenko,  
F. Niedermayer, T. Ortin, S. Pokorski, A. Pomarol, 
F. Quevedo, C. Savoy, M. Shaposhnikov,  and A. Tseytlin  for
useful comments.

This work was supported in part by DOE under the grant number
DE-FG02-94ER40823.

\vspace{0.3cm}

{\bf Note Added}: \hspace{0.2cm} After this work was submitted to hep-th
our attention was drawn to some recent results in $M$-theory
which, conceptually, overlap with some ideas discussed above. A construction
suggested in Ref. \cite{Hor}, in the context of $E_8\times E_8$
heterotic string, treats the degrees of freedom of our four-dimensional world
(below some threshold energy) as those living on a boundary of a 
five-dimensional manifold. The bulk
five-dimensional supergravity serves as a messenger of SUSY breaking.
We are grateful to  P. Ho\v{r}ava for his kind communication.

A. Kusenko has pointed out to us that one can attempt to use the ideas
presented in this paper to put supersymmetry on the lattice,
in the same vein as Kaplan's program puts chiral fermions on the
lattice.

\end{document}